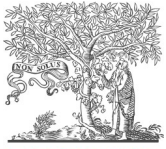
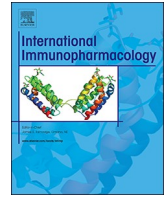

# Use of Machine Learning and Artificial Intelligence to predict SARS-CoV-2 infection from Full Blood Counts in a population

Abhirup Banerjee[a,1], Surajit Ray[b,1], Bart Vorselaars[c,1], Joanne Kitson[d], Michail Mamalakis[e], Simonne Weeks[f], Mark Baker[g], Louise S. Mackenzie[f,⁎]

[a] *Department of Engineering Science, Institute of Biomedical Engineering, University of Oxford, OX3 7DQ, UK*
[b] *School of Mathematics and Statistics, University of Glasgow, Glasgow G12 8QW, UK*
[c] *School of Mathematics and Physics, University of Lincoln, Brayford Pool, Lincoln LN6 7TS, UK*
[d] *School of Computer Science, Electrical and Electronic Engineering, and Engineering Maths, University of Bristol, Merchant Venturers Building, Woodland Rd, Clifton, Bristol BS8 1UB, UK*
[e] *School of Computer Science, University of Sheffield, 211 Portobello, Sheffield City Centre, Sheffield S1 4DP, UK*
[f] *School of Pharmacy and Biomedical Sciences, University of Brighton, BN2 4GJ, UK*
[g] *Hypatia Solutions Ltd, Impact Hub King's Cross, 34b York Way, King's Cross, London, XGL N1, UK*

ABSTRACT

Since December 2019 the novel coronavirus SARS-CoV-2 has been identified as the cause of the pandemic COVID-19. Early symptoms overlap with other common conditions such as common cold and Influenza, making early screening and diagnosis are crucial goals for health practitioners. The aim of the study was to use machine learning (ML), an artificial neural network (ANN) and a simple statistical test to identify SARS-CoV-2 positive patients from full blood counts without knowledge of symptoms or history of the individuals. The dataset included in the analysis and training contains anonymized full blood counts results from patients seen at the Hospital Israelita Albert Einstein, at São Paulo, Brazil, and who had samples collected to perform the SARS-CoV-2 rt-PCR test during a visit to the hospital. Patient data was anonymised by the hospital, clinical data was standardized to have a mean of zero and a unit standard deviation. This data was made public with the aim to allow researchers to develop ways to enable the hospital to rapidly predict and potentially identify SARS-CoV-2 positive patients.

We find that with full blood counts random forest, shallow learning and a flexible ANN model predict SARS-CoV-2 patients with high accuracy between populations on regular wards (AUC = 94–95%) and those not admitted to hospital or in the community (AUC = 80–86%). Here, AUC is the Area Under the receiver operating characteristics Curve and a measure for model performance. Moreover, a simple linear combination of 4 blood counts can be used to have an AUC of 85% for patients within the community. The normalised data of different blood parameters from SARS-CoV-2 positive patients exhibit a decrease in platelets, leukocytes, eosinophils, basophils and lymphocytes, and an increase in monocytes.

SARS-CoV-2 positive patients exhibit a characteristic immune response profile pattern and changes in different parameters measured in the full blood count that are detected from simple and rapid blood tests. While symptoms at an early stage of infection are known to overlap with other common conditions, parameters of the full blood counts can be analysed to distinguish the viral type at an earlier stage than current rt-PCR tests for SARS-CoV-2 allow at present. This new methodology has potential to greatly improve initial screening for patients where PCR based diagnostic tools are limited.

## 1. Introduction

The World Health Organization (WHO) characterized the COVID-19 pandemic on 11th March 2020 [1]. The symptoms of COVID-19 induced by the novel pathogen SARS-CoV-2, are difficult to differentiate from other common infections in a large proportion of those infected. It is estimated that about 40% of cases will experience mild disease (cough, fever), 40% experience moderate disease (bilateral pneumonia), 15% develop severe disease and 5% will have critical disease [2]. A recommendation by the WHO has been to conduct early screening of patients to identify, isolate and track those infected and prevent transmission [2].

Reverse transcription Polymerase Chain Reaction (rt-PCR) based methodologies have been the gold standard in confirming that the individual presenting with COVID-19 has active viral shedding of SARS-CoV-2 [3,4]. However, the ability to conduct wide scale testing of patients has been limited by a number of factors including suitable resources for rt-PCR based testing for the presence of SARS-CoV-2. In addition, the standard test used has an 80% accuracy (compared to chest CT scan results) [5], which may depend on the specific level of viral shedding by any individual at the time of sample test. Moreover, the time from sample collection to informing the patient is estimated to take many hours to several days according to the systems in place. These complex issues hand in hand with the wide-ranging symptoms presenting in patients and the discrepant results between chest CT, symptoms and rt-PCR results [5], indicates that testing for the direct presence of virus requires significant improvement.

While highly specific tests for SARS-CoV-2 are under development using CRISPR [6] and Biosensors [7,8], these innovative applications will require highly specialised equipment and resources. This will negatively affect less affluent areas that will be unable to access these technologies in the short time frame in order to contain the pandemic. Therefore, there is a global challenge to enable screening without the need for sophisticated equipment and resources.

Machine learning (ML) and artificial intelligence (AI) approaches are rapidly being developed to aid clinical procedures in the current pandemic, such as predicting the specificity of new therapies [9] and diagnosing COVID 19 patients from radiographic patterns on CT scans [10].

In the current study we focused on predicting whether a person is SARS-CoV-2 positive or negative in the early stage of the disease. The approach taken is based on the reported limitation to conduct rt- PCR tests and the need to quickly differentiate between individuals presenting with similar symptoms as seen between COVID 19 and other common infections. The dataset used here comes from a public challenge by mindstream-ai [11] to use AI to predict the test result for SARS-CoV-2 (positive/negative) solely from full blood counts.

The Hospital Israelita Albert Einstein is located in the state of Sao Paulo, Brazil, with a population of 12 million people had 477 confirmed cases of Covid 19 and 30 associated deaths by the 23rd March 2020. The hospital publicly released full blood counts [12] from 598 anonymised patients, along with the rt-PCR result for SARS-CoV-2 and their age quantile (symptoms or patient history were not released). Here we present evidence that patients having SARS-CoV-2 can be identified by random forest, ML and artificial neural networks (ANN) to patients not admitted to hospital (community) and to patients in a regular ward setting through recognition of the altered immune cell profile. These will allow for a rapid early screening purely based on the blood profile.

## 2. Methods

### 2.1. Patient dataset

All data processed in this study is published on a public forum [12], as part of a challenge to accelerate approaches to combat the spread of SARS-CoV-2. This dataset contains anonymized data from patients seen at the Hospital Israelita Albert Einstein, at São Paulo, Brazil, and who had samples collected to perform the SARS-CoV-2 rt-PCR and additional laboratory tests during a visit to the hospital. All data were anonymized following the best international practices and recommendations [11]. All clinical data were standardized to have a mean of zero and a unit standard deviation.

**Table 1**
Sao Paulo Dataset groups; only datasets from patients with full blood counts and rt-PCR SARS-CoV-2 outcome included; Pathogen test conducted on 356 subset of the 598 tested for SARS-CoV-2.

| Number of patients | Community | Regular Ward | Semi Intensive Unit | Intensive Care Unit ICU) | Total |
|---|---|---|---|---|---|
| **SARS-CoV-2 negative** | 431 (92%) | 31 (54%) | 34 (81%) | 21 (72%) | 517 (86%) |
| **SARS-CoV-2 positive** | 39 (8%) | 26 (46%) | 8 (19%) | 8 (28%) | 81 (14%) |
| **Diagnosis of other pathogens (non SARS-CoV-2)** | 149 (32%) | 12 (21%) | 17 (40%) | 12 (41%) | 188 (31%) |

Data provided included age (percentile group), outcome from rt-PCR SARS-CoV-2 test and standard full blood count: hematocrit, haemoglobin, platelets, mean platelet volume (MPV), red blood cells (RBC), lymphocytes, mean corpuscular haemoglobin concentration (MCHC), leukocytes, basophils, neutrophils, mean corpuscular haemoglobin (MCH), eosinophils, mean corpuscular volume (MCV), monocytes and red blood cell distribution width (RBCDW) [13].

The full dataset released included 5,644 individual patients tested between the 28th of March 2020 and 3rd of April 2020, of which 598 full blood count results were used for statistical analysis. The remaining 5046 results were not used due to lack of full blood count data.

Without in-depth data on the timeline of the tests performed in the duration of the infection, analysis was performed on the basis of severity according to the location of the patient within the hospital system. The blood counts are from four classifications of patients: Community (patients not admitted to hospital), Regular Ward, Semi Intensive Unit and Intensive Care Unit (ICU) (Table 1).

Patients that are in semi-intensive unit and ICU were excluded from our modelling, so as to ensure prediction is based on early indicators. Also neutrophils are not included, since this is not reported for all 598 patients. Furthermore, we exclude age from our modelling.

There is a large imbalance of positive (8%) vs negative (92%) SARS-CoV-2 patients in the community. As a result it is more informative to test separately for the specificity (% of negative patients correctly identified as negative) and the sensitivity (% of positive patients correctly identified as positive) rather than solely the total accuracy.

Of the 598 complete counts analysed, 367 patients had also been tested for other pathogens: Adenovirus, *Bordetella pertussis*, *Chlamydophila pneumoniae*, Coronavirus 229E, Coronavirus HKU1, Coronavirus NL63, Coronavirus OC43, Influenza A H1N1 2009, Influenza A, Influenza B, Metapneumovirus, Parainfluenza 1, Parainfluenza 2, Parainfluenza 3, Parainfluenza 4, Respiratory Syncytial Virus or Rhinovirus Enterovirus.

### 2.2. Model definitions: random forest and Lasso based regularized generalized linear models and artificial neural network

For our 2-class (SARS-CoV-2 positive vs negative) classification we employ several ML and ANN network models.

For the ML models we apply random forest [14–16] and Lasso-elastic-net regularized generalized linear (glmnet) models for classification. Random forest as a classifier is based on several decision trees. To classify a new object, each decision tree provides a classification for input data and the random forest uses the mode of those classification to decide on the class.

In this paper, glmnet, on the other hand, fits a traditional logistic model. But instead of using all the predictors, it uses a regularized path to select the most important variables and only use them in the logistic regression. For both these methods, we present results for 10-fold cross-validation and their corresponding variable importance plots [17].

Furthermore, an ANN [13] model is defined with 14 input parameters and three hidden layers and trained for 100 epochs. The classification performance of the ANN model is evaluated with stratified 10-fold cross validation.

### 2.3. Model performance measures

The performance of each model is expressed in terms of AUC, sensitivity, specificity and accuracy. They are defined as follows: sensitivity is the fraction of the SARS-CoV-2 positive patients correctly identified; specificity is the fraction of SARS-CoV-2 negative patients correctly identified as such; accuracy is the total number of patients correctly identified. By lowering the threshold of detecting SARS-CoV-2 positive patients, the sensitivity can increase at the expense of specificity. Hence we also look at the commonly employed AUC. This is the area under the receiver-operating characteristics curve; the curve when plotting sensitivity vs (1-specificity) upon changing the threshold. The AUC, also known as the Wilcoxon-Mann-Whitney statistic, is the probability that a SARS-CoV-2 positive patient is higher ranked than a SARS-CoV-2 negative patient. A higher AUC generally implies a better performing model. We note that the drawback of using accuracy alone is that, in an unbalanced set of mainly SARS-CoV-2 negative patients (as is the case in the community dataset), the accuracy can be high using zero sensitivity.

## 3. Results

### 3.1. Statistical analysis

Significant differences ($p < 0.05$) between 9 of the 15 blood count parameters were shown between patients in a regular ward setting who tested positive or negative to SARS-CoV-2 presence (Fig. 1). In order of importance (lower $p$) for the significant increased values for SARS-CoV-2 positive patients: MPV > RBC > lymphocytes > hematocrit > hemoglobin. The decreased values are eosinophils > leukocytes > platelets. For community-patients we found statistically significant differences ($p < 0.05$) in 6 blood count parameters; the increased values are monocytes > MPV, while leukocytes > platelets > neutrophils > eosinophils show a significant decrease for SARS-CoV-2 positive patients.

### 3.2. Modelling

Among the 598 patients with full blood count profiles, 57 were admitted in the regular ward (26 tested rt-PCR positive to SARS-CoV-2, and 31 negative). Furthermore, a total of 470 patients were not admitted to the hospital (39 tested positive for SARS-CoV-2 and 431 negative). We will report model predictions for both sets of patients.

The defined ANN model for the regular ward patients produces an average classification accuracy of 90% over stratified 10-fold cross-validation. The receiver operating characteristic (ROC) curves for all 10 folds and the values of area under the curve (AUC) are presented in Fig. 2a, along with the average AUC (0.95 ± 0.08). The normalized confusion matrix corresponding to fold 2 (the worst AUC value), is presented in Fig. 2b.

The defined ANN model for the community ward produces an average classification accuracy of 89% over stratified 10-fold cross-validation. The ROC curves for all 10-folds produce an average AUC of 0.77 ± 0.08, while the sensitivity and specificity indices are estimated as 0.28 and 0.95, respectively. Since the imbalance between two classes in the dataset (positive/negative = 0.09) degrades the performance of the defined ANN model, the Synthetic Minority Oversampling Technique (SMOTE) [18] is adapted for balancing the two classes in the training dataset. The stratified 10-fold cross-validation technique is again incorporated and it is repeated for 10 times. The average accuracy, AUC, sensitivity, and specificity are estimated as 0.87, 0.80, 0.43, and 0.91, respectively. The ROC curves and the values of AUC for one repetition of the stratified 10-fold cross-validation are presented in Fig. 3a, along with the average AUC (0.80 ± 0.05). The normalized confusion matrix, corresponding to fold 8, is presented in Fig. 3b.

The results from the implementation of random forest and glmnet on the 57 patients in the regular ward gives an average AUC of 94% over 10 fold classification, while for the patients in the community the AUC is 84–86%. The full array of sensitivity, specificity and accuracy values for optimal choices of cutoff for these two classifiers are given Table 2 and Table 3. As compared to the ANN, both random forest and glmnet provide more insight into the most important variables (Fig. 4(a) and (b)) and a clear indication of how the decision has been obtained. Additionally, glmnet does a variable selection, by providing a much smaller stable set of variables among 14 highly correlated predictors.

Analysis of the top four variables according to glmnet that corresponded to patients in community (Supplementary Fig. 1) and regular ward (Supplementary Fig. 2) indicated a clear recognisable pattern. In particular to note in patients in the regular ward is in the decreased pool of leukocytes overall, increase in red blood cells, and in particular a significant decrease in eosinophils. Community patients having SARS-CoV-2 have distinctively high levels of monocytes and low levels of leukocytes.

In order to enable a rapid prediction model for clinics [12] where clinicians may want to choose only two (Fig. 5a) three (Fig. 5b) or four parameters (Fig. 5c), the monocytes/leukocytes/eosinophils/platelets trends in Supplementary Figs. 1 and 2 and Fig. 1 were analysed by adding monocytes and subtracting leukocytes, eosinophils and platelets. This difference indicates little overlap between patients who test positive and negative for SARS-CoV-2, however it must be noted that this is a simple additive formula based on normalised data.

For example a simple logistic regression (LR) with derived variable $y =$ monocytes - leukocytes - eosinophils - platelets, shows this blood characteristic can predict the SARS-CoV-2 test outcome with an average AUC = 85% over 10 fold cross-validation among the patients in the community, and AUC = 81% for patients in regular ward.

The model predictions from ANN, random forest, glmnet and the formula: monocytes - leukocytes - eosinophils (and m-l-e-p) are summarized in Table 2 and Table 3.

Of the patients testing for SARS-CoV-2 and full blood count tests, 366 patients were tested for other pathogens, of which 188 patients were diagnosed with other infections, to note Rhinovirus and Influenza B (Table 4). Collectively, 51% tested positive for: Respiratory Syncytial Virus, Influenza A, Influenza B, Parainfluenza 1, Coronavirus NL63, Rhinovirus Enterovirus, Coronavirus HKU1, Parainfluenza 3, *Chlamydophila pneumonia*, Adenovirus, Parainfluenza 4, Coronavirus 229E, Coronavirus OC43, Influenza A H1N1 2009, *Bordetella pertussis*, Metapneumovirus and Parainfluenza 2. The similar presentation of many of these infections to COVID19 may however be differentiated by the clear difference in immune response to SARS-CoV-2 using the ANN model of full blood count analysis.

Of the 598 patients, only one tested positive for SARS-CoV-2 and for at least one other pathogen; that patient also tested positive for Influenza B and for Coronavirus NL63 and was in ICU.

In addition to the changing profile of immune cells, a change in red blood cells and platelets were noted. In order to describe the profile of cells, the mean changes were plotted from patients in Regular Ward (Supplementary Fig. 3).

## 4. Discussion

We developed multiple independent models (statistical, random forest and shallow learning) that can predict SARS-CoV-2 with an AUC of up to 86% for community and 95% for regular ward patients, using only data collected from their normalized full blood counts. This provides an initial screen of SARS-CoV-2 positive from negative using biomarkers at an early stage in the disease presentation. This screen has been conducted on a set of data based on severity judged by the location of the patient in hospital (admitted to the regular ward compared to not admitted to hospital; ICU patients were excluded). Hence the models are able to distinguish from altered blood profiles in patients who were later diagnosed with other pathogens.

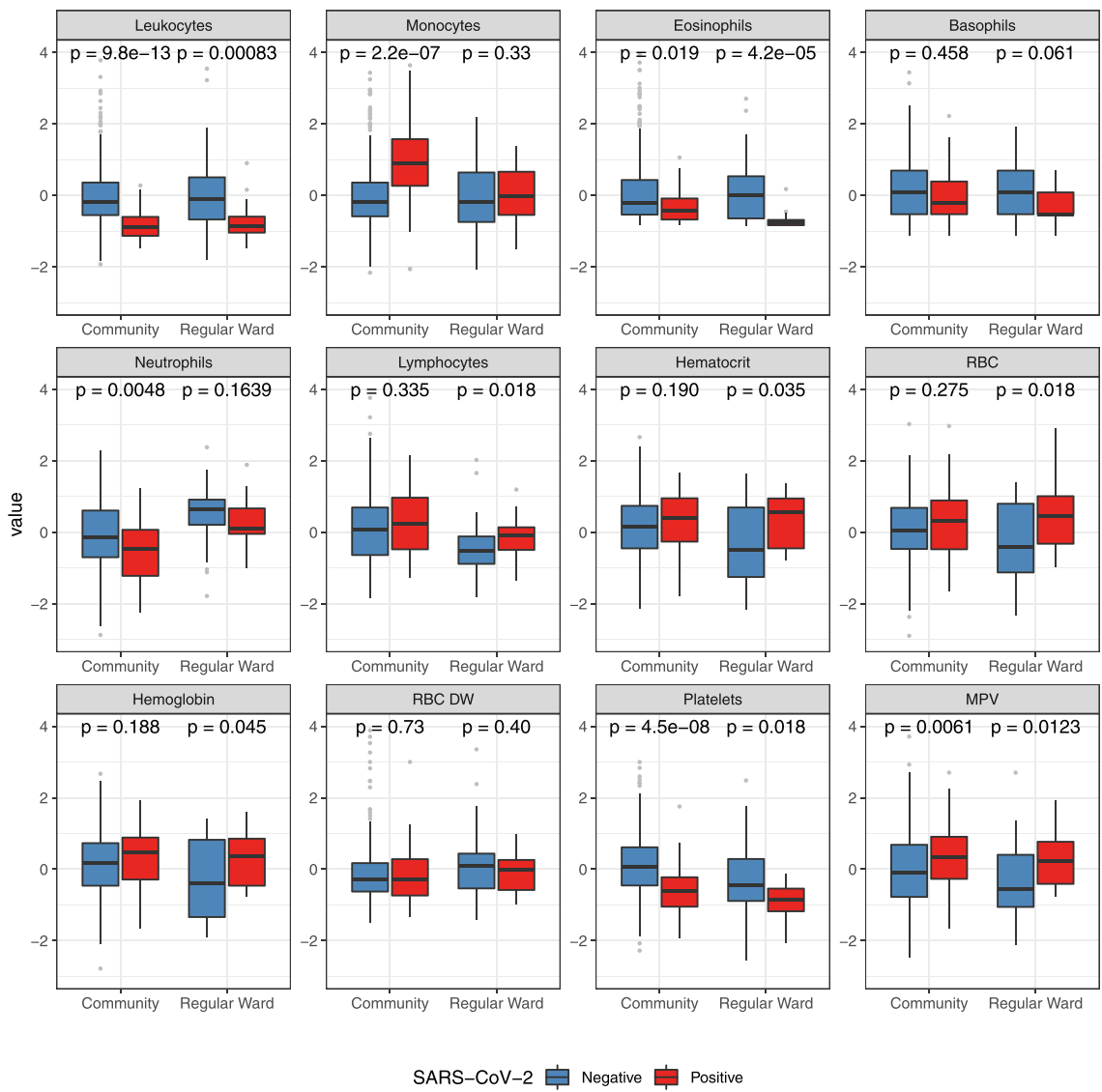

**Fig. 1.** Box plots showing median and 1st/3rd quartile of individual parameters of the full blood counts categorized by whether tested positive (red box) or negative (blue box) by the rt-PCR test for SARS-CoV-2 and by whether they remained in the community or were admitted in the regular ward. MPV; mean platelet volume, RBC; red blood cells, RBCDW; red blood cell distribution width. The $p$-values are tests of equality of population using the Wilcoxon rank sum test, where $p < 0.05$ implies statistically significant difference between the populations. (For interpretation of the references to colour in this figure legend, the reader is referred to the web version of this article.)

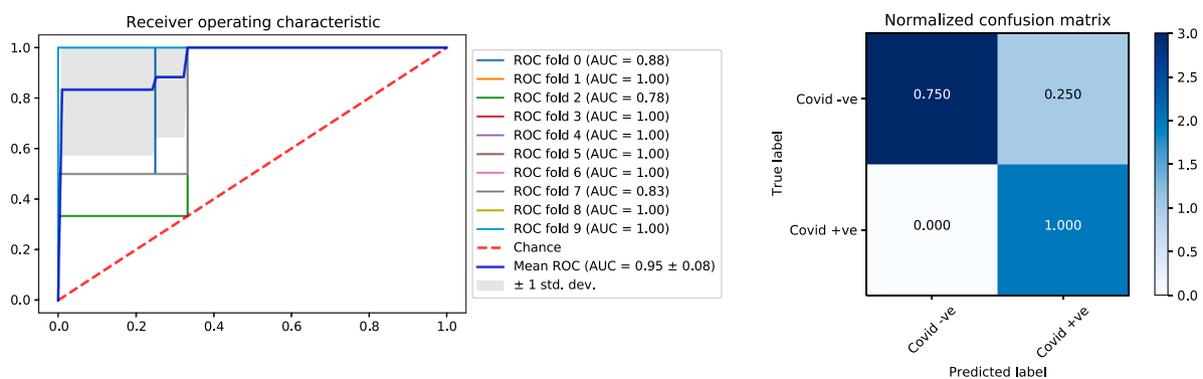

**Fig. 2. a.** The ROC curve of the defined ANN model over patients admitted to regular ward; **b.** The normalized confusion matrix, corresponding to fold 2 (corresponds to the worst value of AUC). ROC: Receiver operating characteristic; AUC: area under the curve.

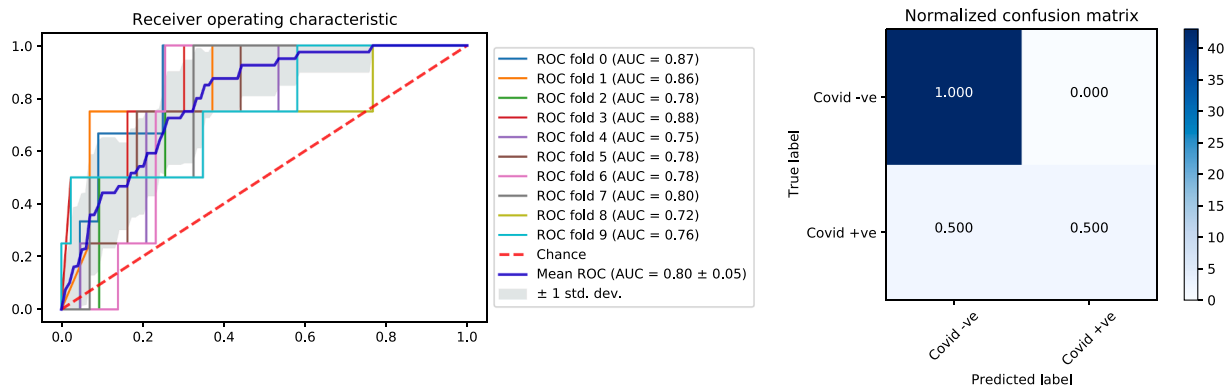

**Fig. 3. a.** The ROC curve of the defined ANN model over patients not admitted to the hospital (community); **b.** The normalized confusion matrix, corresponding to fold 8 (corresponds to the worst value of AUC). ROC: Receiver operating characteristic; AUC: area under the curve.

**Table 2**
Model predictions for patients in regular ward testing for SARS-CoV-2 positive.

| Variables | Classifier | Sensitivity | Specificity | Accuracy | AUC |
| --- | --- | --- | --- | --- | --- |
| 14 different blood counts | ANN | 0.85 | 0.94 | 0.90 | 0.95 |
| 14 different blood counts | RF | 0.82 | 0.90 | 0.91 | 0.94 |
| 14 different blood counts | glmnet | 0.92 | 0.93 | 0.91 | 0.94 |
| y = monocytes - leukocytes - eosinophils - platelets | LR | 0.85 | 0.77 | 0.81 | 0.81 |
| y = monocytes - leukocytes - eosinophils | LR | 0.85 | 0.71 | 0.77 | 0.79 |
| y = monocytes - leukocytes | LR | 0.65 | 0.58 | 0.61 | 0.65 |

**Table 3**
Model predictions for patients not admitted to hospital (community) testing for SARS-CoV-2 positive.

| Variables | Classifier | Sensitivity | Specificity | Accuracy | AUC |
| --- | --- | --- | --- | --- | --- |
| 14 different blood counts | ANN | 0.43 | 0.91 | 0.87 | 0.80 |
| 14 different blood counts | RF | 0.60 | 0.88 | 0.82 | 0.86 |
| 14 different blood counts | glmnet | 0.65 | 0.81 | 0.81 | 0.84 |
| y = monocytes - leukocytes - eosinophils - platelets | LR | 0.82 | 0.78 | 0.79 | 0.85 |
| y = monocytes - leukocytes - eosinophils | LR | 0.82 | 0.79 | 0.79 | 0.84 |
| y = monocytes - leukocytes | LR | 0.74 | 0.77 | 0.77 | 0.81 |

It is well recognised that the symptoms of COVID-19 are accompanied by a significant change in immune response [19], with a decreased population of leukocytes, lymphocytes [20,21] and eosinophils [19–21] found throughout all stages of infection. Indeed it was suggested in one case report from Wuhan that eosinopenia together with lymphopenia may be a potential indicator for diagnosis [22]. Other early reports used similar predictive patterns found in Full Blood Count parameters, suggesting that elevated neutrophil to lymphocyte ratio could be used as part of the diagnosis [23]. In that study they investigated the change in blood parameters in a total of 93 patients (severe and non severe collected together), using commonly used ratios used in the diagnosis of viral respiratory diseases such as: neutrophil-to-lymphocyte ratio, derived NLR ratio (d-NLR, neutrophil count divided by the result of WBC count minus neutrophil count), platelet-to-lymphocyte ratio and lymphocyte-to-monocyte ratio. However, these parameters commonly alter in respect to other viral infections, and we initiated research to identify new ratios to distinguish SARS-CoV-2 from other pathogens.

Here we show that a new simple calculation based on leukocytes, monocytes, eosinophils and platelets (normalized monocytes - leukocytes - eosinophils - platelets) can be used to predict with 85% AUC the presence of SARS-CoV-2 for early-stage community patients. Further validation will be required to determine whether our model can distinguish fully from other pathogens, although initial small numbers indicate a trend that is positive.

Leukocytes are a family of white blood cells that includes neutrophils, lymphocytes, monocytes, eosinophils and basophils. Specific pathogens induce specific responses, such as an increase in circulating neutrophils and monocytes typically increase during sepsis as part of the immune defence mechanism, and several studies have indicated that ratios between different blood cells can be used to predict outcome. The ratio of monocyte: neutrophil can be used to determine sepsis severity [24], and platelet: monocyte aggregation has been reported to be increased in patients with Influenza A (H1N1) [25]. Indeed, the clinical interest in using the relationship between different parameters of the full blood counts is of great value and interest due to its simplicity and readily measurable parameters. While not yet in common clinical use, it has been shown that neutrophil: lymphocyte and mean platelet volume: platelet count could predict the outcome for critically ill patients with peritonitis and pancreatitis (bacteremia) [26]. Our results presented here clearly demonstrates that an early prediction of those infected with SARS-CoV-2 may be made by the relationship between monocytes, eosinophils and leukocytes to differentiate them from other pathogen induced infections.

The data analysed in this report focussed on the clear early stages in order to differentiate between patients presenting with common symptoms prior to the need to be in ICU. The changes in parameters of the full blood counts is easily distinguishable from other pathogen-induced infections, with an apparent decrease in leukocyte population consisting of a large proportion of monocytes. This observation is in keeping with other tools used to diagnose the severity of COVID-19 by the measurement of IL6 which significantly increases throughout the progression of disease. The cytokine IL6 is predominantly synthesised by monocytes and macrophage (which derive from monocytes), and is partly responsible for the drive of the immune response to deleterious cytokine storm.

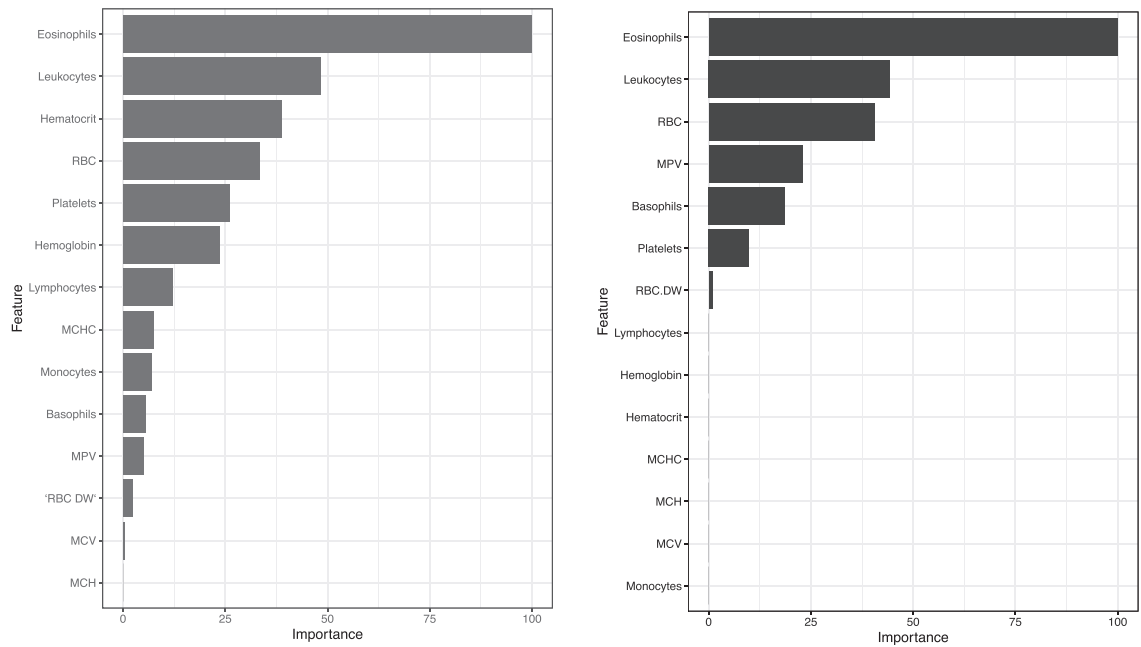

**Fig. 4.** Variable importance plot of (a) random forest and (b) glmnet classification of SARS-CoV-2 positive patients who are in the regular hospital ward. The plot shows the importance of variables in building the respective predictive model. MPV; mean platelet volume, RBC; red blood cells, MCHC; mean corpuscular haemoglobin concentration, MCH; mean corpuscular haemoglobin, MCV; mean corpuscular volume, RBCDW; red blood cell distribution width. (For interpretation of the references to colour in this figure legend, the reader is referred to the web version of this article.)

In order to best understand the immune response to SARS-CoV-2 it is useful to compare it to other similar coronaviruses. There are two main example: closest known example is that of SARS-CoV virus (SARS) which infected 8,096 people and killed 774 in 2002–2003, and Middle East Respiratory syndrome coronavirus (MERS-CoV) which in 2013 infected 2,102 people and killed 780. The majority of all other coronaviruses infect the upper respiratory tract and cause mild respiratory and gastrointestinal infections. In contrast, the highly infectious and highly pathogenic coronaviruses SARS and MERS were noted to have similar effects on the immune response, and more recently overlaps with the pathogenesis now witnessed with SARS-CoV-2 infections. The SARS virus directly infects human monocytes, which then produces the cytokines that attract neutrophils, macrophage and activated T lymphocytes [27]; MERS was shown to increase monocytes and their IL6 production [28]. In parallel, this study indicates that the pathogenesis to SARS-CoV-2 may be linked to monocytes and the production of IL6. Here in our study, the analysis of monocyte involvement will be crucial in the prediction of SARS-CoV-2 infection. Additionally, use of AI and ML to recognise the altered pattern of key blood parameters will be a useful tool in future with the emergence of any future coronavirus that is equally pathogenic and contagious.

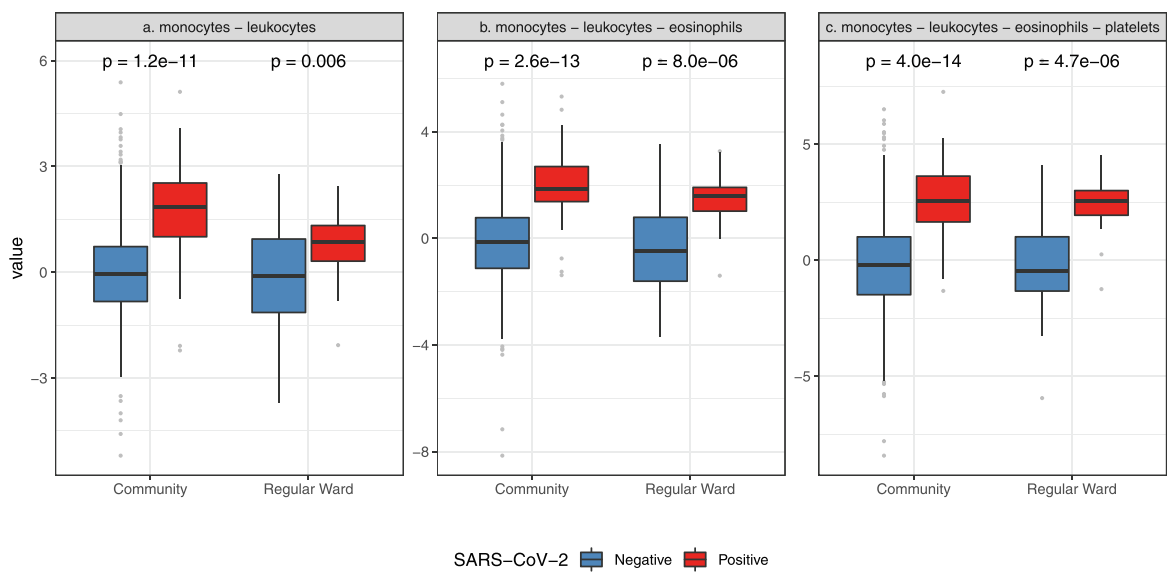

**Fig. 5.** Box plots of blood characteristics: **a.** monocytes - leukocytes (*m-l*) **b.** monocytes - leukocytes - eosinophils (*m-l-e*) and **c.** monocytes - leukocytes - eosinophils - platelets (*m-l-e-p*); all normalized values, categorized by whether tested negative (blue box) or positive (red box) to rt-PCR SARS-CoV-2 test, and whether they remained in the community or were admitted in the regular ward. All *p*-values are tests of equality of population using Wilcoxon rank sum test and suggest statistically significant difference. (For interpretation of the references to colour in this figure legend, the reader is referred to the web version of this article.)

**Table 4**
Sao Paulo Dataset: table showing number of patients who tested positive for pathogens other than SARS-CoV-2, by patient admission (all patients had been tested for SARS-CoV-2 and had full blood count results).

| Pathogen | Community | Regular ward | Semi Intensive Unit | ICU | Total per pathogen |
|---|---|---|---|---|---|
| Adenovirus | 0 | 0 | 0 | 0 | 0 |
| *Bordetella pertussis* | 18 | 1 | 1 | 1 | 21 |
| *Chlamydophila pneumoniae* | 1 | 1 | 1 | 0 | 3 |
| Coronavirus 229E | 1 | 0 | 0 | 0 | 1 |
| Coronavirus HKU1 | 0 | 0 | 0 | 0 | 0 |
| Coronavirus NL63 | 9 | 1 | 1 | 1 | 12 |
| Coronavirus OC43 | 3 | 0 | 0 | 0 | 3 |
| Influenza A H1N1 2009 | 2 | 0 | 0 | 0 | 2 |
| Influenza A | 3 | 0 | 0 | 0 | 3 |
| Influenza B | 20 | 2 | 2 | 1 | 25 |
| Metapneumovirus | 0 | 0 | 0 | 0 | 0 |
| Parainfluenza 1 | 0 | 0 | 0 | 0 | 0 |
| Parainfluenza 2 | 4 | 1 | 0 | 0 | 5 |
| Parainfluenza 3 | 4 | 0 | 1 | 0 | 5 |
| Parainfluenza 4 | 1 | 0 | 0 | 0 | 1 |
| Respiratory Syncytial Virus | 3 | 0 | 2 | 4 | 9 |
| Rhinovirus Enterovirus | 78 | 6 | 9 | 5 | 98 |
| **Total** | **147** | **12** | **17** | **12** | **188** |

The decrease in platelets in patients testing positive for SARS-CoV-2 in the regular ward is opposite to that observed in separate reports of patients with Influenza A [29]. A subset of the patients included in this analysis were tested for other pathogens, and a proportion of those negative for SARS-CoV-2 were positive for other pathogens including Coronaviruses (NL63, HKU1 and 229E) and few had Influenza A or B. This suggests that platelet count is a good indicator in this predictive model, and may be a good way to differentiate SARS-CoV-2 infection from Influenza A.

The suggests a potential decreased platelet presence in patients, which is of concern due to its link to thrombocytopenia and increased internal bleeding. This supports other reports of an increase in thrombocytopenia being associated with higher mortality in COVID-19 patients [21]. The condition of thrombocytopenia may be the reason for the recently noted rashes observed in patients, especially young children [30]. However there have been reports that clotting is increased in COVID-19 patients, and the data we show is normalised and may be misleading. This normalised data also indicates an increase in platelet size (MPV), which suggests that there is rapid platelet production by the bone marrow. Indeed, it is recommended that COVID-19 patients are administered antiplatelet therapy to protect against thrombosis [31]. Overall this highlights the difficulty of interpreting mean normalised data.

This report is the first to use primary patient data of full blood counts to test and train an ANN to predict from patients in a regular ward as well as those in the community who will test positive for SARS-CoV-2. This preliminary model will be further trained and adapted with the aim to address the shortfall in direct SARS-CoV-2 testing methods in hospitals. This will enable a prediction that allows health care providers to conduct rapid cheap screening to separate patients into those who are most likely to have SARS-CoV-2 and those who do not. Early screening allows segregation of patients and early treatment intervention.

**Funding**

This research did not receive any specific grant from funding agencies in the public, commercial, or not-for-profit sectors.

**Acknowledgements**

Shelley Taylor, Taylormade Business Consultants for project managing the team during and after the Corona hackathon competitions.

**Appendix A. Supplementary material**

Supplementary data to this article can be found online at https://doi.org/10.1016/j.intimp.2020.106705.

**References**

[1] WHO Virtual press conference full transcript, https://www.who.int/docs/default-source/coronaviruse/transcripts/who-audio-emergencies-coronavirus-press-conference-full-and-final-11mar2020.pdf (accessed 24/04/2020).
[2] WHO Covid-19 Strategy Update, https://www.who.int/docs/default-source/coronaviruse/covid-strategy-update-14april2020.pdf (accessed 24/04/2020).
[3] Ying-Hui Jin, Lin Cai, Zhen-Shun Cheng, Hong Cheng, Tong Deng, Yi-Pin Fan, Cheng Fang, Di Huang, Lu-Qi Huang, Qiao Huang, Yong Han, Bo Hu, Fen Hu, Bing-Hui Li, Yi-Rong Li, Ke Liang, Li-Kai Lin, Li-Sha Luo, Jing Ma, Lin-Lu Ma, Zhi-Yong Peng, Yun-Bao Pan, Zhen-Yu Pan, Xue-Qun Ren, Hui-Min Sun, Ying Wang, Yun-Yun Wang, Hong Weng, Chao-Jie Wei, Dong-Fang Wu, Jian Xia, Yong Xiong, Hai-Bo Xu, Xiao-Mei Yao, Yu-Feng Yuan, Tai-Sheng Ye, Xiao-Chun Zhang, Ying-Wen Zhang, Yin-Gao Zhang, Hua-Min Zhang, Yan Zhao, Ming-Juan Zhao, Hao Zi, Xian-Tao Zeng, Yong-Yan Wang, Xing-Huan Wang, A rapid advice guideline for the diagnosis and treatment of 2019 novel coronavirus (2019-nCoV) infected pneumonia (standard version), Military Med. Res. 7 (1) (2020), https://doi.org/10.1186/s40779-020-0233-6.
[4] K.H. Hong, S.W. Lee, T.S. Kim, H.J. Huh, J. Lee, S.Y. Kim, J.S. Park, G.J. Kim, H. Sung, K.H. Roh, J.S. Kim, H.S. Kim, S.T. Lee, M.W. Seong, N. Ryoo, H. Lee, K.C. Kwon, C.K. Yoo, Guidelines for laboratory diagnosis of coronavirus disease 2019 (COVID-19) in Korea, Ann. Lab. Med. 40 (5) (2020) 351–360.
[5] G. Lippi, A.M. Simundic, M. Plebani, Potential preanalytical and analytical vulnerabilities in the laboratory diagnosis of coronavirus disease 2019 (COVID-19), Clin. Chem. Lab. Med. (2020).
[6] J.P. Broughton, X. Deng, G. Yu, C.L. Fasching, V. Servellita, J. Singh, X. Miao, J.A. Streithorst, A. Granados, A. Sotomayor-Gonzalez, K. Zorn, A. Gopez, E. Hsu, W. Gu, S. Miller, C.Y. Pan, H. Guevara, D.A. Wadford, J.S. Chen, C.Y. Chiu, CRISPR-Cas12-based detection of SARS-CoV-2, Nat. Biotechnol. (2020).
[7] G. Qiu, Z. Gai, Y. Tao, J. Schmitt, G.A. Kullak-Ublick, J. Wang, Dual-functional plasmonic photothermal biosensors for highly accurate severe acute respiratory syndrome coronavirus 2 detection, ACS Nano (2020).
[8] G. Seo, G. Lee, M.J. Kim, S.H. Baek, M. Choi, K.B. Ku, C.S. Lee, S. Jun, D. Park, H.G. Kim, S.J. Kim, J.O. Lee, B.T. Kim, E.C. Park, S.I. Kim, Rapid detection of COVID-19 causative virus (SARS-CoV-2) in human nasopharyngeal swab specimens using field-effect transistor-based biosensor, ACS Nano (2020).
[9] B.R. Beck, B. Shin, Y. Choi, S. Park, K. Kang, Predicting commercially available antiviral drugs that may act on the novel coronavirus (SARS-CoV-2) through a drug-target interaction deep learning model, Comput. Struct. Biotechnol. J. 18 (2020) 784–790.
[10] C. Butt, G. J., D. Chun, B.A. Babu, Deep learning system to screen coronavirus disease 2019 pneumonia, Appl. Intell. (2020) 1–7.
[11] Mindstream-ai CoronaHack - AI vs Covid-19 https://www.coronahack.co.uk/ (accessed 14/04/2020).
[12] Data4u, E. Hospital Israelita Albert Einstein, Sao Paulo, Brazil, Diagnosis of Covid-19 and its clinical spectrum, 3/2020, https://www.kaggle.com/einsteindata4u/covid19 (accessed 14/04/2020).
[13] X. Troussard, S. Vol, E. Cornet, V. Bardet, J.P. Couaillac, C. Fossat, J.C. Luce, E. Maldonado, V. Siguret, J. Tichet, O. Lantieri, J. Corberand, French-Speaking Cellular Hematology, G., Full blood count normal reference values for adults in France, J. Clin. Pathol. 67 (4) (2014) 341–344.
[14] L. Breiman, Random forests, Machine Learning 45 (2001) 5–32.
[15] M.L. Calle, V. Urrea, Letter to the editor: stability of Random Forest importance measures, Brief Bioinform. 12 (1) (2011) 86–89.
[16] H. Wang, F. Yang, Z. Luo, An experimental study of the intrinsic stability of random forest variable importance measures, BMC Bioinformatics 17 (2016) 60.

[17] A. Fisher, C. Rudin, F. Dominici, All models are wrong but many are useful: variable importance for black-box, proprietary, or misspecified prediction models, using model class reliance, Mathematics (2018).

[18] N. Chawla, K.W. Bowyer, L.O. Hall, P.W. Kegelmeyer, SMOTE: synthetic minority over-sampling technique, J. Artif. Intell. Res. 16 (2002) 3210357.

[19] C. Qin, L. Zhou, Z. Hu, S. Zhang, S. Yang, Y. Tao, C. Xie, K. Ma, K. Shang, W. Wang, D.S. Tian, Dysregulation of immune response in patients with COVID-19 in Wuhan, China, Clin. Infect. Dis. (2020).

[20] H. Yun, Z. Sun, J. Wu, A. Tang, M. Hu, Z. Xiang, Laboratory data analysis of novel coronavirus (COVID-19) screening in 2510 patients, Clin. Chim. Acta (2020).

[21] X. Yang, Q. Yang, Y. Wang, Y. Wu, J. Xu, Y. Yu, Y. Shang, Thrombocytopenia and its association with mortality in patients with COVID-19, J. Thromb. Haemost. (2020).

[22] J.J. Zhang, X. Dong, Y.Y. Cao, Y.D. Yuan, Y.B. Yang, Y.Q. Yan, C.A. Akdis, Y.D. Gao, Clinical characteristics of 140 patients infected with SARS-CoV-2 in Wuhan China, Allergy (2020).

[23] A.P. Yang, J.P. Liu, W.Q. Tao, H.M. Li, The diagnostic and predictive role of NLR, d-NLR and PLR in COVID-19 patients, Int. Immunopharmacol. 84 (2020) 106504.

[24] W.F. Fang, Y.M. Chen, Y.H. Wang, C.H. Huang, K.Y. Hung, Y.T. Fang, Y.C. Chang, C.Y. Lin, Y.T. Chang, H.C. Chen, K.T. Huang, Y.C. Chen, C.C. Wang, M.C. Lin, Incorporation of dynamic segmented neutrophil-to-monocyte ratio with leukocyte count for sepsis risk stratification, Sci. Rep. 9 (1) (2019) 19756.

[25] M.T. Rondina, B. Brewster, C.K. Grissom, G.A. Zimmerman, D.H. Kastendieck, E.S. Harris, A.S. Weyrich, In vivo platelet activation in critically ill patients with primary 2009 influenza A(H1N1), Chest 141 (6) (2012) 1490–1495.

[26] D. Djordjevic, G. Rondovic, M. Surbatovic, I. Stanojevic, I. Udovicic, T. Andjelic, S. Zeba, S. Milosavljevic, N. Stankovic, D. Abazovic, J. Jevdjic, D. Vojvodic, Neutrophil-to-lymphocyte ratio, monocyte-to-lymphocyte ratio, platelet-to-lymphocyte ratio, and mean platelet volume-to-platelet count ratio as biomarkers in critically ill and injured patients: which ratio to choose to predict outcome and nature of bacteremia? Mediators Inflamm. 2018 (2018) 3758068.

[27] W. Hu, Y.T. Yen, S. Singh, C.L. Kao, B.A. Wu-Hsieh, SARS-CoV regulates immune function-related gene expression in human monocytic cells, Viral Immunol. 25 (4) (2012) 277–288.

[28] C.K. Min, S. Cheon, N.Y. Ha, K.M. Sohn, Y. Kim, A. Aigerim, H.M. Shin, J.Y. Choi, K.S. Inn, J.H. Kim, J.Y. Moon, M.S. Choi, N.H. Cho, Y.S. Kim, Comparative and kinetic analysis of viral shedding and immunological responses in MERS patients representing a broad spectrum of disease severity, Sci. Rep. 6 (2016) 25359.

[29] M. Koupenova, H.A. Corkrey, O. Vitseva, G. Manni, C.J. Pang, L. Clancy, C. Yao, J. Rade, D. Levy, J.P. Wang, R.W. Finberg, E.A. Kurt-Jones, J.E. Freedman, The role of platelets in mediating a response to human influenza infection, Nat. Commun. 10 (1) (2019) 1780.

[30] J.D. Bouaziz, T. Duong, M. Jachiet, C. Velter, P. Lestang, C. Cassius, A. Arsouze, E. Domergue Than Trong, M. Bagot, E. Begon, L. Sulimovic, M. Rybojad, Vascular skin symptoms in COVID-19: a French observational study, J. Eur. Acad. Dermatol. Venereol. (2020).

[31] Jing-Chun Song, Gang Wang, Wei Zhang, Yang Zhang, Wei-Qin Li, Zhou Zhou, Chinese expert consensus on diagnosis and treatment of coagulation dysfunction in COVID-19, Military Med. Res. 7 (1) (2020), https://doi.org/10.1186/s40779-020-00247-7.